# Respiratory Sound Classification Using Long-Short Term Memory

Chelsea Villanueva[1]
cvillanueva@scu.edu

Joshua Vincent[2]
jvincent@scu.edu

Alexander Slowinski[1]
aslowikowski@scu.edu

Mohammad-Parsa Hosseini[2]
phosseini@scu.edu

1. Department of Computer Engineering, Santa Clara University, Santa Clara, CA, 95118
2. Department of Bioengineering, Santa Clara University, Santa Clara, CA, 95118

*Abstract—* **Developing a reliable sound detection and recognition system offers many benefits and has many useful applications in different industries. This paper examines the difficulties that exist when attempting to perform sound classification as it relates to respiratory disease classification. Some methods which have been employed such as independent component analysis and blind source separation are examined. Finally, an examination on the use of deep learning and long short-term memory networks is performed in order to identify how such a task can be implemented.**

*Keywords—sound classification, deep learning, LSTM, machine learning*

## I. Introduction

Developing a reliable sound detection and recognition system offers many benefits and has many useful applications in different industries. There have been many efforts made in the process of developing these reliable sound detection systems. These efforts include the incorporation of machine learning and neural networks using deep learning [27-30]. Many different algorithms and filters have been used in an attempt to increase accuracy, reliability, and repeatability in different environments [37-40]. There are many challenges posed to developers in designing such systems. In order to produce a system that is reliable and accurate, many variables must be taken into consideration due to the distorting effects they may have on sound waves. Some of the variables that must be addressed include the effects of background noise, weather, echos, and the effects of the surrounding architecture on sound waves. In developing a reliable sound classification system, these variables must be addressed.

This paper seeks to examine the different methods that have been used in the detection and classification of sounds as well as their effectiveness, specifically through the use of deep learning and machine learning techniques. Machine learning and deep learning is often used for sound classification systems. Additionally, we will examine some of the techniques used to minimize negative effects of environmental variables. Lessons learned from these previous attempts are applied in order to enhance current neural network sound classification methods for specific use in sound classification related to respiratory disease classification. Finally, we will show an example of using the long short-time memory (LSTM) network to classify audio files relating to respiratory health.

Machine learning is a method of learning through the use of algorithms and inputs. A computer processes input data in order to make a decision and also has the ability to learn. Deep learning is a field within machine learning and is modeled after the learning and decision making processes of the human brain. Deep learning involves the use of two or more hidden layers which consist of neurons. The neural network is trained through the use of a supervised or unsupervised learning process. There are many different types of neural network structures within deep learning. The structure which works best depends on the type of application and data being processed. The learning process involves input data traversing the neural network of hidden layers which contain neurons which are all linked together. A neuron is linked with the other neurons in the preceding and successive layers, with each neuron processing the data based on a function, known as an activation function which modifies the input values.

After making a pass through the network, a prediction is made based on the input values and a loss is also calculated so that the network can be trained based on how far off the prediction was from the expected output. This loss, or cost function, is used to adjust the weights throughout the layers of the network [5]. One such method is known as gradient descent and is used in combination with backpropagation. With backpropagation, the weights of the neurons are adjusted backwards from the output toward the front of the network by adjusting the loss with respect to the weights. Using backpropagation, the weights are able to be adjusted in one pass. Using stochastic gradient descent, the neurons are adjusted at random.

By running through multiple iterations, the network is able to learn and its accuracy will increase. Adjustments to the network may also be made to better the rate of learning or to increase accuracy; some of these adjustments include which activation functions are used and at which point in the network, the number of layers, and width of the layers. Deep learning is a kind of machine learning that is able to learn using unsupervised feature extraction [31-36]. Unsupervised networks differ from supervised in that they do not need to be trained by a "professional." In other words, supervised networks are initially trained using labeled data to concretely specify the outputs of the data inputs. Using a trained dataset, the network learns to differentiate different types of labeled data. With unsupervised networks, unlabeled datasets are used and the network learns to identify differences amongst the data by extracting features from the data and making different connections and recognizing patterns amongst the data. Some examples of machine learning include speech recognition which can be found in products such as Apple Siri, the Google Home Assistant, and Amazon Alexa. In this paper, we will make an examination on the use of long short-term memory and how it is used as a deep learning approach for sound classification purposes.

We applied a simple LSTM model on a dataset of audio files that capture the respiratory sounds of a number of healthy and unhealthy patients. Classification of certain characteristics found in these audio files will indicate a certain disease and can be used as an automated tool for diagnosis. Distinguished sounds like 'wheezes' and 'crackles' are usually the classes for classification that indicate a certain respiratory disease. There are certain features that differentiate crackles from wheezes, such as the length of sound, continuity, pitch, etc. [19]. This is a typical classification done by multiple models, and so the model proposed is novel because it goes even further, by using the actual respiratory diseases as classes. With this deep learning network, doctors and patients can be more confident in a diagnosis.

## II. LITERATURE REVIEW

Sound classification as it relates to the positive detection and classification in gunshots can be broken down into different goals. One foreseeable problem involves the use of microphone sensors in an outdoor public environment. With the sensors being exposed to different combinations of sounds, the network should be able to isolate and differentiate different sounds. There have been many methods proposed and it also gave rise to the famous problem known as the cocktail party. A network can be set up in a way that it is able to isolate individual sounds, such as muting or minimizing all but one of the voices from a group of people talking. This application can also be extended to this sensor network where similar sounding noises should be identified and separated.

In an indoor and outdoor environment, the same sound tends to reach the same point more than once due to the architecture of the surrounding areas giving rise to echos. These echos can be understood as a recurrent phenomenon and are seen as a convolutional mixture in sound recordings [1]. These echo sounds effects should be removed in order to optimize sound analysis and [11, 1]. This mixture of sounds can be represented by the equation:

$$\mathbf{x}(n) = \sum_{k=0}^{K-1} \mathbf{A}_k \mathbf{s}(n-k) + \boldsymbol{\nu}(n). \quad (1)$$

Independent Component Analysis (ICA) which attempts to separate the sound signals into independent components based on their source. ICA has been one approach developed to mitigate this phenomenon [1, 11]. This analysis can take on different approaches, such as the one by Barros where the closest detected sound source signal is the one which is isolated. In our application, however, this would not be useful as we may end up removing the desired sounds from being analyzed. Instead, the approach which should be taken is to isolate the different sounds from the environment into individual signals, an approach made by Makino [7].

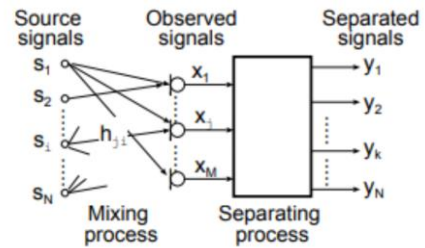

Figure 1. A block diagram of an undetermined BSS [7]

Another issue which causes difficulty in separating sound signals is that not only are the number of sound source signals unknown, but they are also greater than the number of microphone sensors analysing the data. Blind Source Separation (BSS) is the method which will be most productive given our requirements. Using an array of microphones, Makino et al. were able to perform ICA by detecting the sparseness of a sound signal between the microphone arrays. This was accomplished by evaluating the zero point sound signals in order to identify individual signals and combining it with the musical mixture created by the sounds [2, 8].

Before developing these sound classification models, data input is required, however, data related to these abnormal events is limited. In order to develop a large dataset useful for training neural networks, one method which has been proposed is the use of data augmentation. Data augmentation consists of using already present data inputs and augmenting them slightly so that they can be used as additional inputs, greatly increasing the dataset which can be used for training. Some of the data augmentation which can be used involve either slowing or speeding up the audio sample, known as time stretching, raising or lowering the pitch of the audio sample, mixing the audio sample with another sound containing background noise, and compressing the dynamic range of the audio sample [12].

Another approach is to use cross acoustic transfer learning [7]. One observation made from analyzing the spectrograms of speech and sounds is that when enlarged, they show similar acoustic patterns. By using speech audio samples, the dataset for training sounds can be greatly increased, making the neural network more reliable. Once the data set can be improved with these methods, the type of network needs to be chosen.

A similar pattern of classification emerges from various works related to respiratory health and deep learning [41]. Many works use the 2017 International Conference on Biomedical Health Informatics Challenge (ICBHI) data set, which contains audio files regarding respiratory health. Three papers discuss using this data set to train and test models to classify wheezes and crackles from normal sound. Therefore, it is valuable to compare these works and how the models produce different results. To differentiate crackles and wheezes, the model must train based on features such as frequency, length, and pitch [17]. The purpose of classifying sounds is to utilize this information to make easier diagnoses, since a specific sound can be indicative of a certain disease [15].

Many related works use deep learning to approach respiratory health in this way, however, they differ in the network architectures they use. Chen et al. had the most optimal results, by using an optimized s-transform (OST) to process a spectrogram out of the raw sound files, and then using a deep residual network (ResNet) to train and test the model [15]. This model was able to attain a 98% accuracy, 96% sensitivity, and a 100% specificity [15]. Although these are very promising results, Chen et al. does not dive into a technical explanation of the actual model, which makes it useless for our purposes in creating our own model.

The other two works by Ma et al. and Perna et al. involve different network architectures including a bi-ResNet and a recurrent neural network, both of which do not perform as well as the former method [16, 17]. However, they both pose similar goals to the first, where the classification is based on wheeze, crackle, or neither. For our implementation we wanted to classify something more unique. In order to have clearer diagnoses we have decided to classify based on a number of diseases, rather than just the labeling of a sound. This is a unique classification as compared to other works.

There has been previous work done using LSTM models for respiratory issues, but most have been applied to different types of datasets, other than audio files. Many related works revolve around spectrograms, i.e. visual graphs of audio files. For example, one paper converts a dataset of audio files recording people using inhalers into spectrograms, before normalizing and running the data into the model. It was successful in accuracy, with a 92-94% accuracy, and quite good at differentiating background noise from the target data [18]. With a similar process, LSTM also worked on signal image data for sleep therapy [20]. The LSTM model is ideal when there is data that corresponds to time because there may be certain periods of importance, whereas other periods may actually make results worse [20]. LSTM networks can learn to optimize certain periods of time in order to train for the desired classifications.

There are many optimizations of LSTM that can also be implemented, and will be discussed further in section III. Hajiaghayi et al. utilizes LSTM for pattern recognition in recognizing singular events that can either cause or block code failure. The purpose of this paper is unrelated to our goals, however, their unique use of the LSTM network could be one way to approach our own issue. In order to gain insights into their sequential data, they take bits and pieces of their data out of the testing circuit until they find a change in a result/prediction. Those retracted data sections are of importance, and therefore can be recognized in other sequences [21]. We could produce the same retraction method to our audio files for recognition of key data points.

Kumar et al. shows that LSTM models need not be complicated with many layers. The purpose of their network is to predict any trends of depression from a data set of EEG waveform charts. Their network consists of 1 LSTM layer, 10 hidden neurons, dropout, and an input and output. It predicts the testing data with only .006% loss. Utilizing the fundamentals of LSTM, i.e adjusting for the forget gate, weights, and bias, can produce results that are very promising, especially considering it is a basic network [22]. This would be the type of network that we would want to construct when considering our short time frame and knowing that we have a limited amount of computing power.

Ren et al. combines two different networks, which are recurrent neural networks: LSTM and the Attention Mechanism. The bi-directional LSTM layer and the attention layer work hand in hand as they both end up outputting sentence features for its question and answer classification. It was successful, which shows the possibilities of combining two methods together [23]. Cai et al. goes even further in combining two different categories of neural networks: a convolutional neural network and recurrent neural network. With a similar task as Ren et al., they are trying to classify different pairs of embeddings of query based data. Its accuracy ranges from 88 to 97%, while Ren et al. only receives an accuracy of 85 to 95%. They both do very well and better than in comparison to a lone LSTM network [24]. With a more complicated architecture and varied layers, they are able to change a lot more parameters in accordance to their goal. This would definitely be an option to explore with our own data set, for future endeavors.

Kumar et al. uses the same CNN and LSTM combined approach when trying to detect falls in elderly homes. However, what is emphasized in this work is the efficiency of power and computation. With the goal of the system architecture being able to be commercialized, it was important to note that this type of network could produce accurate results and still be affordable to reproduce [25]. The commercial aspect and energy consumption data of this work resonates with the goals of our own system.

III. METHODS

LSTM is a type of recurrent neural network which addresses the shortfalls of long term dependencies that exist with traditional RNNs. With traditional RNNs, large gaps between references of data can lead to vanishing gradients which cause the network to learn slowly or not when stochastic gradient descent is used with the backpropagation optimization algorithm. LSTM uses a recurrent architecture to pass a cell state and the hidden state in a chain-like fashion; the data passed between cells is modified slightly through the use of pointwise operations. LSTM addresses the shortfalls of RNN with the addition of forget gates, input gates, and output gates.

The forget gate uses a sigmoidal activation function on the input cell state and hidden state from the previous cell to output a vector with values ranging from 0 to 1; a 0 meaning the cell can completely forget the data, and a 1 meaning to remember all of the data. The input gate also applies a sigmoidal activation function, as well as a hyperbolic tangent activation function on the previous hidden state and cell state to identify what data should be added to the cell state. The output gate also uses the sigmoidal and hyperbolic tangent activation functions to identify which data to send as output and which data to send to the next cell within the hidden state. The following image depicts an LSTM neuron [3]:

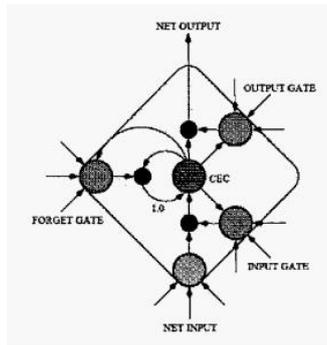

Figure 2. The architecture of a basic LSTM neuron [3]

LSTM has benefits when used with time series data, which includes sound, especially with a sound classification system because, as previously mentioned, RNNs are susceptible to a vanishing gradient and large gaps between sound input sources may lead to the network's inability to accurately train or classify data [5, 6]. The following formulas represent those used with LSTM [14]:

$$i_t = \sigma(W_i([x_t, y_{t-1}])) \quad (2)$$
$$f_t = \sigma(W_f([x_t, y_{t-1}])) \quad (3)$$
$$o_t = \sigma(W_o([x_t, y_{t-1}])) \quad (4)$$
$$g_t = \tanh(W_g([x_t, y_{t-1}])) \quad (5)$$
$$c_t = f \odot c_{t-1} + i \odot g \quad (6)$$
$$y_t = o \odot \tanh(c_t) \quad (7)$$

Some of the sound classification methods which have been proposed are modeled after speech recognition methods and use deep learning methods. In using deep learning, long short term memory appears to be most suitable with sound analysis because sound uses time series data. In order to classify sounds using this methods, features are extracted from the sound inputs using the mel-frequency cepstral coefficients (MFCCs), zero crossing rate, harmonic coefficients, LPCs, and PLPs [6]. MFCC is a method used to represent the short term power spectrum of sound and is widely used in sound feature extraction [3, 6, 14]. Following feature extraction, similar features are grouped together using a Gaussian mixture model [9, 10].

A common theme among researchers in attempting to develop a sound classification model which seeks to identify "abnormal" sound events when developing an application which attempts to identify stressful or potentially violent situations is that the model must be able to differentiate background noise from foreground noise [9, 10]. One method in addressing this issue is to subtract what the estimated background noise would be from the input sound so that only the audio event remains. Another approach is to use a controlled environment with known background noise to train the model, which requires a large training dataset [9, 10].

One modification on the LSTM model has been to use a bidirectional LSTM (BLSTM) which has shown great promise in improving the prediction and accuracy rate with feature extraction [4, 14]. In the following example by Graves, bi-directional LSTM was shown to more accurately predict the features of the target data than a forward only net or a reverse only net:

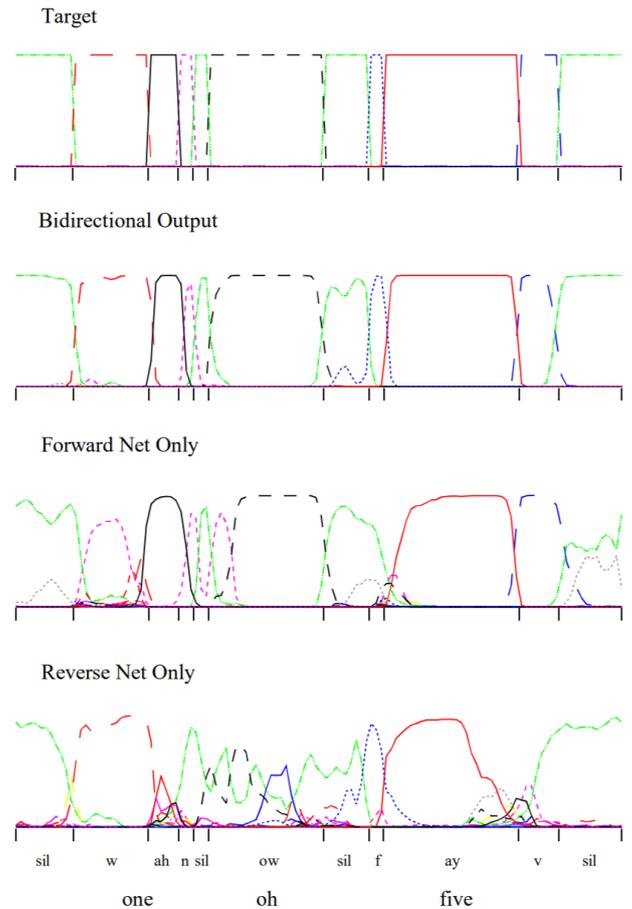

Figure 3. A comparison of bidirectional LSTM with forward net only and reverse net only feature extraction. This example shows how bidirectional LSTM was able to make more accurate depictions which closely resembles the target data [4]

BLSTM simply uses two LSTMs, one for the forward and one for the reverse processing of the data and then integrating their output using the following formula [4, 14]:

$$\boldsymbol{y}(t) = \boldsymbol{y}_F(t) \oplus \boldsymbol{y}_B(N - t + 1) \qquad (8)$$

## IV. DATA SET

The data set we use for classification of the respiratory diseases comes together from multiple research sources, including ICBHI, University of Aveiro in Portugal, and Aristotle University of Thessaloniki in Greece. All of the files were recorded with permission from each of the subjects. For each of the patient files, names and personal identifiable information was removed, in order to keep anonymity [19]. This is important for the ethical use of the data set. In total, there are 920 sound files accumulated from 126 patients. The subjects ranged in age from elderly to adult to child [19]. Diversity of the age groups reduces the bias of the results towards one age group. Therefore, the model can be generalized to more patients, instead of over fitting to a specific characterized group.

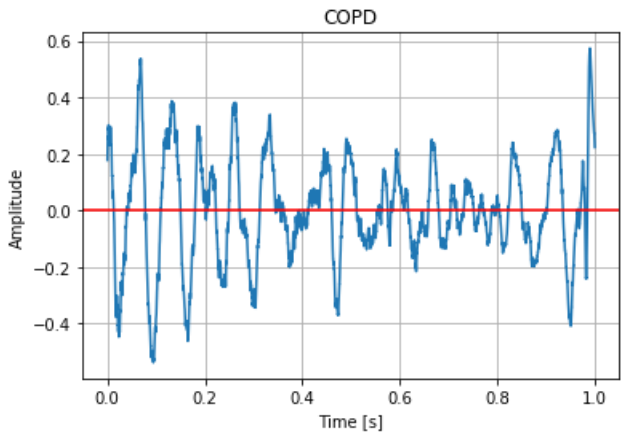

Figure 4: One second sample from the Respiratory Sound Dataset.

The data set is organized in the following for each of the audio files: demographics of the patient, patient number from 1 to 126, the methodology for obtaining the recording, what time in the respiratory cycle is being recorded, the presence of crackles or wheezes, and one of the eight classes of diseases that has been diagnosed. The Respiratory Sounds Data set includes respiratory recordings from 126 separate patients. Doctors diagnosed each patient with one of eight classes: healthy, asthma, COPD, LRTI, URTI, bronchiectasis, bronchiolitis, or pneumonia. Each of these diagnoses were finalized by multiple respiratory experts [19]. The labeling of the diagnoses needs to be trusted. With multiple experts reviewing and diagnosing the patients, there is greater confidence in the validity of this data set. Since the diagnosis of each patient was provided, we could perform supervised learning on the data set.

Prior to training the model, we looked at the distribution of classes within the dataset. As depicted in Figure 5, class 2, which represents COPD, comprises the overwhelming majority of the dataset. We will discuss the implications of this class distribution in section VI.

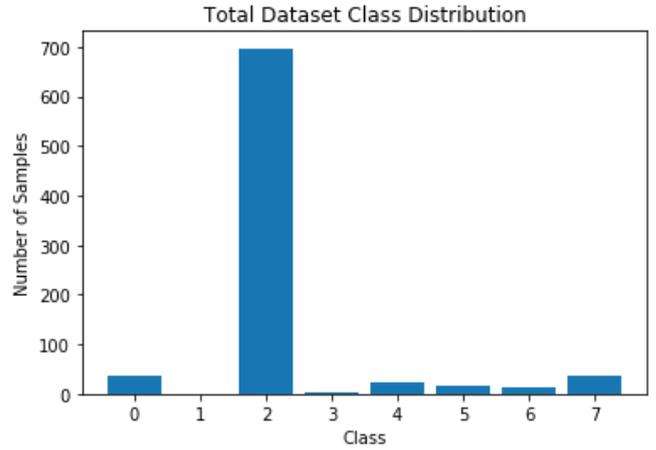

Figure 5: COPD comprises the overwhelming majority of the dataset.

While loading the dataset, we grabbed a one second sample from each audio file. We used one second samples in the interest of time. Using multiple tensorflow libraries we were able to gather and normalize the data samples collected from our data set. In order to input our audio files into an LSTM network we needed to follow certain tutorials in regards to the TFRecordDataset library [26].

## V. TRAINING

We trained a simple LSTM model on the dataset discussed in section IV. The model, itself, consisted of an LSTM layer and a fully connected layer with eight neurons. Each neuron represented one of the eight classes in the dataset. The model architecture, as well as training procedure is available on Github:
ttps://github.com/jvincent144/Respiratory_Sounds.git.

We split the dataset into three subsets: training, validation, and testing. To review, the model learns on the training subset and checks generalizability on the validation subset. The model evaluates its performance on the testing subset, an unseen set of samples, after training. We included 70% of samples in the training set, 10% of samples in the validation set, and 20% in the testing set.

The training and validation losses decreased, as in Figure 5, which indicated that the model learned on the training subset and generalized on the validation subset. In addition, the training and validation accuracies increased to a maximum of 51%, as in Figure 6. We would like to reiterate that the model discriminates between eight classes. Therefore, the probability of randomly selecting the correct class is 12.5%. During the peak of training, this model achieved a 51% accuracy; in other words, the model selected four out of eight classes correctly.

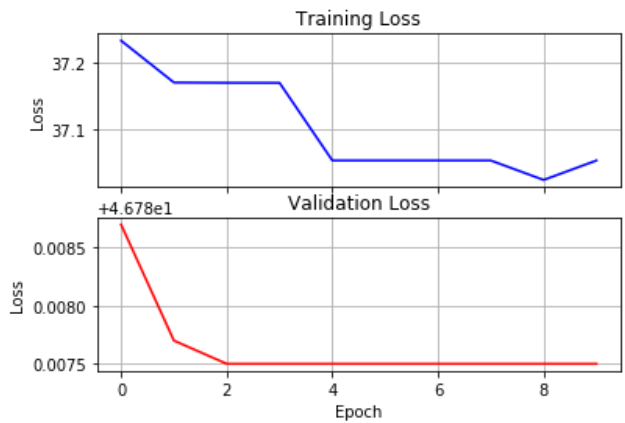

Figure 6: The training loss and validation loss fell. Therefore, the model learned on the training dataset and generalized to the validation dataset.

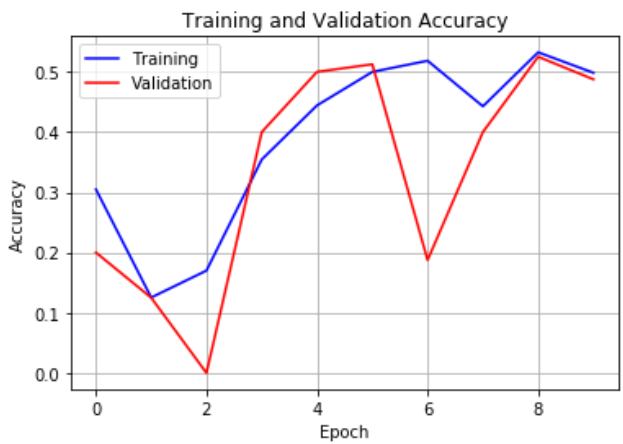

Figure 7: The training accuracy and validation accuracy increased over the course of 10 epochs.

## VI. RESULTS & DISCUSSION

After training, we evaluated the model on the testing subset. The model achieved a 46% accuracy on the testing subset. Therefore, the model predicted approximately four out of eight classes correctly, an improvement from random guessing.

As in Figure 6, the validation loss plateaued during the third epoch of training. Therefore, the model stopped learning generalizable features. We speculate that this plateau is due to the dataset distribution discussed in section IV. In order to prevent this plateau in the future, we may take a couple actions including: (1) augment the dataset and (2) increase sample length. Recall that the dataset is skewed toward COPD patients. The dataset contains a limited number of cases of LRTI, URTI, bronchiectasis, bronchiolitis, and pneumonia. Therefore, the model has a limited number of general cases to learn from. We can increase the sample size of these cases through data augmentation, such as subsampling from the LRTI, URTI, bronchiectasis, bronchiolitis, and pneumonia audio at multiple distinct time frames. In addition, we can add a time shift to increase the number of samples. In addition, we can increase the sample length from one second to four seconds in order to capture an entire deep breath.

The accuracy could have been improved by implementing some of the optimizations as discussed earlier. Makino et al. or Barros et al suggests two different methodologies in separating background noise from the target sounds. These algorithms may have been applied for our own uses regarding crackles and wheezes. Proceeding further, we may also consider using a bi-directional LSTM model, whose accuracy is compared in section III.. Overall, there are many modifications that can improve our basic LSTM network.

## VII. CONCLUSION

Using these different methodologies, we seek to be able to differentiate different sound signal sources and reduce the environmental noise and echo effects present in such a sound detection and classification system. Although not implemented in our own LSTM network, it would make great improvements to our current work. There is a lot of research which has been put forth showing promise in this technology which can be used to create effective public safety systems and diagnosis tools.